\documentclass[twocolumn,english,aps, pra, twocolumn, superscriptaddress, showpacs]{revtex4-2}
\usepackage[T1]{fontenc}
\usepackage[utf8]{inputenc}
\usepackage{color}
\usepackage{amstext}
\usepackage{amssymb}
\usepackage{graphicx}
\usepackage{esint}

\makeatletter
\usepackage{babel}
\usepackage{hyperref}
\hypersetup{
    colorlinks=true,
    linkcolor=red,
    citecolor=blue,
    filecolor=blue,      
    urlcolor=blue,
}
\makeatother

\usepackage{babel}
\begin{document}
\title{A quantum Otto-type heat engine with fixed frequency}
\author{Richard Q. Matos}
\address{Instituto de Fí­sica, Universidade Federal de Goiás, 74.001-970, Goiânia
- GO, Brazil}
\author{Rogério J. de Assis}
\address{Departamento de Fí­sica, Universidade Federal de São Carlos, 13.565-905,
São Carlos - SP, Brazil}
\author{Norton G. de Almeida}
\address{Instituto de Fí­sica, Universidade Federal de Goiás, 74.001-970, Goiânia
- GO, Brazil}
\begin{abstract}
In this work, we analyze an Otto-type cycle operating with a working
substance composed of a quantum harmonic oscillator (QHO). Unlike
other studies in which the work extraction is done by varying the
frequency of the QHO and letting it thermalize with a squeezed reservoir,
here we submit the QHO to a parametric pumping controlled by the squeezing
parameter and let it thermalize with a thermal reservoir. We then
investigate the role of the squeezing parameter in our Otto-type engine
powered by parametric pumping and show that it is possible to
reach the Carnot limit by arbitrarily increasing the squeezing parameter. Notably, for certain squeezing parameters $r$, e.g.  $r=0.4$, the quasi-static Otto limit can be reached even at non-zero power.
We also investigated the role of entropy production in the efficiency
behavior during the unitary strokes, showing that positive (negative)
changes in entropy production correspond to increases (decreases)
in engine efficiency, as expected. Furthermore, we show that under thermal reservoirs
a work extraction process that is more efficient than the Carnot engine
is impossible, regardless of the quantum resource introduced via the Hamiltonian
of the system. 
\end{abstract}
\maketitle

\section{Introduction}

According to the well-known Carnot result \citep{Carnot1872}, the
maximum efficiency of a reversible heat engine is independent of the
working fluid and occurs at null power, that is, executing
the strokes slowly enough so that the entire cycle is reversible.
Since the emergence of the first quantum heat engines
\citep{Scovil1959,Geusic1967}, there have been efforts to
overcome the so-called Carnot's limit. The search for surpassing
this limit is motivated by the existence of resources not available
at the time of Carnot, which can be accessed when considering
both genuinely quantum reservoirs \citep{Li2011,Wright2018,NuBeler2020}
and working substances that store quantum resources such as
coherence and entanglement \citep{Camati2019,Herrera2022}. For this
reason, the result obtained in Ref. \citep{RoBnagel2014}
was justifiably surprising when announcing that a quantum
heat engine operating at null power can reach efficiencies beyond
Carnot efficiency when using a squeezed hot thermal reservoir, the efficiency
going exponentially to unity when increasing the squeezing parameter.
Another result demonstrating the advantage of quantum heat
engines over classical ones when considering squeezed thermal
reservoirs was obtained in Ref. \citep{Assis2021}, where the authors
show that efficiencies close to unity can be achieved even
when the cycles are performed at non-null power. Refs. \citep{Long2015,Klaers2017,Wang2019,Assis2019,Lee2020,Damas2022,Nettersheim2022}
present other studies that explore the effect of quantum reservoirs
on the performance of quantum heat devices.

The use of squeezed thermal reservoirs raises the question of what impact the squeezing resource would have on the efficiency
of a quantum heat engine if, instead of a squeezed thermal reservoir,
a squeezing operation was carried out directly on the working substance.
To answer this question, we investigate in this paper a quantum
Otto-type heat engine whose working substance is a quantum harmonic
oscillator (QHO) that operates between two conventional thermal reservoirs.
Unlike previous studies, we apply the squeezing operation
directly to the working substance rather than to the reservoirs. Thus,
the asymptotic state of the QHO and, in turn, the engine
efficiency will depend on the squeezing parameter. As we will show,
the best efficiency occurs by fixing the frequency of the QHO and
turning on and off the parametric pumping responsible for squeezing.
Thus, in our proposal the frequency of the free QHO is kept fixed, while the new frequency, obtained by diagonalizing the new Hamiltonian when squeezing is turned on, varies with the intensity of the parametric pumping. For this reason, we call the engine we are proposing an Otto-type quantum heat engine, as it mimics a real Otto quantum heat engine. Also, we will show that it is possible
to achieve efficiencies arbitrarily close to Carnot efficiency by
increasing the intensity of the parametric pumping or, equivalently,
the squeezing parameter. As a corollary of the methods employed here, we show, in an alternative
way to that of Ref. \citep{Campisi2014}, that by using thermal baths,
the maximum efficiency of the Otto engine never\textbf{ surpasses} Carnot, no
matter the quantum resources used.

This paper is organized as follows. In Sec. \ref{subsec:II.A} we
present our model, which consists of a QHO performing work through
a squeezing process. In Sec. \ref{subsec:II.B} we describe the quantum
Otto-type cycle strokes. In Section \ref{subsec:II.C} we diagonalize
the QHO when including the pumping term responsible for squeezing
and we show that the new frequency is lower than the frequency of
the original QHO. In Sec. \ref{subsec:II.D} we calculate the work,
heat, and corresponding efficiency in the quantum Otto-type cycle.
Also, we calculate the efficiency for both high and low temperatures
and compare it with previous results, where the QHO was not subjected
to squeezing. In this Section we also show results for the general
case where the unitary strokes are done either adiabatically or irreversibly.
In Sec. \ref{subsec:II.E} we calculate the efficiency at non-null
power considering the unitary strokes, showing that the entropy production
has a direct relation with the decrease of the engine efficiency. Also, we show that under certain squeezing parameters, it is possible to attain the Carnot limit at non-null power, a result also obtained in Ref.\cite{Assis2021}.
In Sec. \ref{subsec:2.F} we prove that it is not possible, under
thermal reservoirs, to use quantum resources introduced via Hamiltonian
to surpass Carnot. Finally, in Section \ref{sec:III} we present our conclusions.

\section{\label{sec:II}The quantum Otto-type heat engine}

\subsection{\label{subsec:II.A}The working substance}

The working substance we are interested in is a QHO with a
fixed frequency that can be subjected to a squeezing operation and
coupled to a thermal reservoir. The Hamiltonian implementing this
system is
\begin{equation}
H_{SHO}=H_{HO}+H_{SQ},
\end{equation}
where
\begin{equation}
H_{HO}=\hbar\omega\left(a^{\dagger}a+\frac{1}{2}\right)\label{eq:2}
\end{equation}
and 
\begin{equation}
H_{SQ}=\hbar\chi\left[\left(a^{\dagger}\right)^{2}-a^{2}\right],\label{eq:3}
\end{equation}
with $a^{\dagger}$ $\left(a\right)$ being the creation (annihilation)
operator and $\chi$ the coupling constant given by the amplitude
of the parametric pumping, which is related to the squeezing
parameter $r$ by the relation $\chi=\left(\omega/2\right)\tanh\left(2r\right)$. This Hamiltonian
can be implemented in several contexts, as for example cavity QED
\citep{Almeida2004,Villas-Boas2003}, trapped ions \citep{Vogel1995},
and circuit quantum electrodynamics \citep{Blais2020}. Also, squeezing has been done in an experimentally controlled manner in \cite{miwa2014expsqueez,Yoshikawa07expsqueez,miyata2014expsqueez}.

\subsection{\label{subsec:II.B}The quantum Otto-type cycle}

The four strokes of the quantum Otto-type cycle that we intend to study are as follows:\\

\noindent \textbf{\textit{(i) Compression stroke.}} The quantum
Otto-type cycle starts with the free QHO thermalized with
a hot thermal reservoir at temperature $T_{h}$. Thus, with the parametric
pumping absent, the QHO starts the cycle in the Gibbs state $\rho_{1}=\text{e}^{-\beta_{h}H_{HO}}/Z_{h}$,
where $\beta_{h}=1/k_{B}T_{h}$ and $Z_{h}=\text{tr}\left(\text{e}^{-\beta_{h}H_{HO}}\right)$.
Then the QHO is isolated from the hot reservoir, and the parametric
pumping is applied, such that the Hamiltonian of the QHO goes
from $H_{HO}$ to $H_{SHO}$. This stroke occurs unitarily such that the thermalized QHO state evolves from
$\rho_{1}$ to $\rho_{2}=U\rho_{1}U^{\dagger}$, where $U$ is the
unitary evolution operator. This stroke has the name compression
stroke because the effective frequency of the QHO decreases,
which implies a compression in the energy gaps. We can see this compression
by diagonalizing the Hamiltonian $H_{SHO}$, which we do in the following subsection.\\

\noindent \textbf{\emph{(ii) Cooling stroke.}} Here, the harmonic
oscillator is placed in thermal contact with a cold reservoir at temperature
$T_{c}$, keeping its Hamiltonian fixed at $H_{SHO}$ until thermalization.
Then, when thermalizing, the QHO reaches the Gibbs state $\rho_{3}=\text{e}^{-\beta_{c}H_{HO}}/Z_{c}$,
with $\beta_{c}=1/k_{B}T_{c}$ and $Z_{c}=\text{tr}\left(\text{e}^{-\beta_{c}H_{SHO}}\right)$.\\

\noindent \textbf{\emph{(iii) Expansion stroke.}} At this
stroke, reversing the compression stroke, the QHO is decoupled from
the cold reservoir and evolves unitarily as its Hamiltonian changes
from $H_{SHO}$ to $H_{HO}$, driving its state from $\rho_{3}$ to
$\rho_{4}=V\rho_{3}V^{\dagger}$, with $V=U^{\dagger}$.\\

\noindent \textbf{\emph{(iv) Heating stroke.}} Finally, the
QHO is left to thermalize with the hot reservoir with fixed Hamiltonian
$H_{HO}$, thus returning to its initial Gibbs state $\rho_{1}$.\\

Next, we show how to analytically obtain the efficiency of the quantum
Otto-type cycle described above after diagonalizing the Hamiltonian
$H_{SHO}$.

\subsection{\label{subsec:II.C}Diagonalizing the Hamiltonian}

To diagonalize the Hamiltonian $H_{SHO}$, we apply a Bogoliubov transformation
defining a new operator $b$ as 
\begin{equation}
b=\mu a+\nu a^{\dagger},
\end{equation}
where $\mu$ and $\nu$ are complex constants that satisfy
the relation $\mu^{2}-\nu^{2}=1$. It is straightforward to show
that $b$ has the same properties as $a$ so it is also an
annihilation operator on a different basis of the Hilbert space $\left\{ \left|n\right\rangle _{b}:n=0,1,\ldots\right\} $. By selecting the constants $\mu$ and $\nu$ such that the Hamiltonian
has no quadratic terms, we can write 
\begin{equation}
H_{SHO}=\hbar\Omega\left(b^{\dagger}b+\frac{1}{2}\right),\label{eq:5}
\end{equation}
with 
\begin{equation}
\Omega=\omega\sqrt{1-\left(\frac{2\chi}{\omega}\right)^{2}}.\label{eq:6}
\end{equation}
As can be seen, Eq. (\ref{eq:5}) has the same form as Eq.
(\ref{eq:2}) but with modified frequency $\Omega$ and number
operator $b^{\dagger}b$. Bearing in mind that $\chi=\left(\omega/2\right)\tanh\left(2r\right)$,
we see from Eq. (\ref{eq:6}) that $\Omega\leq\omega$, which is why
stroke \emph{(i)} was named \emph{compression stroke}.

The diagonalization performed here allows treating the case of a QHO
and subject to parametric pumping in a formally identical way to that
of a free QHO. The fundamental difference is that, after diagonalizing
$H_{SHO}$, thermalization takes place on the new basis $\left\{ \left|n\right\rangle _{b}:n=0,1,\ldots\right\} $,
with $\rho_{3}$ then being rewritten as $\rho_{3}=\text{e}^{-\beta_{c}\hbar\Omega\left(b^{\dagger}b+1/2\right)}/Z_{c}$,
where $Z_{c}=\text{tr}\Bigl[\text{e}^{-\beta_{c}\hbar\Omega\left(b^{\dagger}b+1/2\right)}\Bigr]$.
Now, it is easy to calculate the efficiency of our quantum heat engine, since the calculation is similar to that made when the working substance is a free QHO with a varying frequency \cite{Abah2012}.
In the following subsection, we calculate the work and heat in the
four strokes described above, which compose the efficiency of the
quantum Otto-type heat engine powered by squeezing.

\subsection{\label{subsec:II.D}Work, heat, and efficiency}

To calculate work $W$ and heat $Q$ at each stroke described
above, we consider the following definitions (see Ref. \citep{Alicki1979}):
$W=\int_{0}^{\tau}dt\text{tr}\bigl[\dot{H}\left(t\right)\rho\left(t\right)\bigr]$
and $Q=\int_{0}^{\tau}dt\text{tr}\left[H\left(t\right)\dot{\rho}\left(t\right)\right]$, where $\tau$ is the duration of the stroke, and $H\left(t\right)$
and $\rho\left(t\right)$ are, respectively, the Hamiltonian and the
state of the working substance at time $t$. According to these definitions,
it is easy to see that there is only work in the expansion and compression
strokes ($W_{exp}$ and $W_{comp}$, respectively), whereas there
is only heat in the heating and cooling strokes ($Q_{h}$ and $Q_{c}$,
respectively). Thus, with the information provided by the previous
subsections, we have 
\begin{equation}
W_{exp}=\frac{\hbar}{2}\left(Q^{*}\omega-\Omega\right)\coth\left(\frac{\beta_{c}\hbar\Omega}{2}\right),\label{eq:7}
\end{equation}
\begin{equation}
W_{comp}=-\frac{\hbar}{2}\left(\omega-Q^{*}\Omega\right)\coth\left(\frac{\beta_{h}\hbar\omega}{2}\right),\label{eq:8}
\end{equation}
\begin{equation}
Q_{h}=\frac{\hbar\omega}{2}\left[\coth\left(\frac{\beta_{h}\hbar\omega}{2}\right)-Q^{*}\coth\left(\frac{\beta_{c}\hbar\Omega}{2}\right)\right],\label{eq:9}
\end{equation}
and
\begin{equation}
Q_{c}=-\frac{\hbar\Omega}{2}\left[Q^{*}\coth\left(\frac{\beta_{h}\hbar\omega}{2}\right)-\coth\left(\frac{\beta_{c}\hbar\Omega}{2}\right)\right],\label{eq:10}
\end{equation}
in which $Q^{*}$ is the Husimi adiabatic parameter
\citep{Husimi1953} that characterizes the speed of the expansion and compression strokes: for a quasi-static (adiabatic)
stroke, $Q^{*}=1$, becoming faster as $Q^{*}$ becomes
bigger than $1$.

With the work and heat at each stroke, we can now obtain the
efficiency of the quantum Otto-type heat engine. The engine efficiency
is given by $\eta=-W_{net}/Q_{abs}$, where $W_{net}=W_{exp}+W_{comp}$
is the net work, the work effectively extracted from the heat engine
($W_{net}<0$), and $Q_{abs}=Q_{h}$ is the heat absorbed by the heat
engine ($Q_{abs}>0$). It is easy to show from Eqs. (\ref{eq:7})-(\ref{eq:10})
that the engine condition $W_{net}<0$  implies $Q_{h}>0$ and $Q_{c}<0$,
which justifies the equality $Q_{abs}=Q_{h}$. Finally, Eqs. (\ref{eq:7})-(\ref{eq:9})
then lead to the engine efficiency 
\begin{equation}
\eta=1-\frac{\Omega}{\omega}\mathcal{F},\label{eq:11}
\end{equation}
where
\begin{equation}
\mathcal{F}=\frac{\coth\left(\frac{\beta_{c}\hbar\Omega}{2}\right)-Q^{*}\coth\left(\frac{\beta_{h}\hbar\omega}{2}\right)}{Q^{*}\coth\left(\frac{\beta_{c}\hbar\Omega}{2}\right)-\coth\left(\frac{\beta_{h}\hbar\omega}{2}\right)}.\label{eq:12}
\end{equation}
If $Q^{*}=1$, Eq. (\ref{eq:11}) takes the simple form 
\begin{equation}
\eta_{QS}=1-\frac{\Omega}{\omega}=1-{\cosh\left(2r\right)},\label{eq:13}
\end{equation}
where the subscript $QS$ stands for quasi-static.

Looking at Eq. (\ref{eq:13}), one might think that $\eta_{QS}=1$
can be reached by taking the limit $r\rightarrow\infty$,
or equivalently $\chi\rightarrow\omega/2$ in Eq. (\ref{eq:6}),
thus eventually surpassing the Carnot efficiency $\eta_{Carnot}=1-\beta_{h}/\beta_{c}$.
However, the engine
condition $W_{net}<0$ provides a limit for the value of the squeezing parameter, $r_{max}=\left(1/2\right)\cosh^{-1}\left(\beta_{c}/\beta_{h}\right)$,
in such a way that $\eta_{QS}\leq\eta_{Carnot}$. In Subsec.
\ref{subsec:2.F}, we show that efficiencies greater than
$\eta_{Carnot}$ are not possible in a very general manner.

It is interesting to compare Eq. (\ref{eq:11}) with the
efficiency of the quantum Otto heat engine of Ref. \citep{Abah2012},
in which a free QHO performs work by modifying its frequency from
$\omega_{2}$ to $\omega_{1}<\omega_{2}$, and whose efficiency is given by \citep{Abah2012}
\begin{equation}
\eta_{Otto}=1-\frac{\omega_{1}}{\omega_{2}}.
\end{equation}

Note that, strictly speaking, if the frequency
of the QHO does not change, the efficiency of the Otto engine is zero ($\eta_{Otto}=0$).
However, considering the redefined frequency $\Omega$ when the Hamiltonian displayed in
Eq. (\ref{eq:5}) is diagonalized, it is possible to obtain the engine
condition taking into account the modified frequency $\Omega$ and,
consequently, the efficiency of the Otto cycle even for fixed frequencies
$\omega$. Also, note that since $\Omega=\omega\sqrt{1-\left(\frac{2\chi}{\omega}\right)^{2}}$ and $\eta_{QS}=1-\frac{\Omega}{\omega}$, if after the parametric pumping was turned
off the QHO frequency had changed to $\omega_{2}>\omega_{1}$, then
$\omega_{1}/\Omega>\omega_{1}/\omega_{2}$. This means that the efficiency
of a quantum Otto cycle including parametric pumping with variable
frequency is always lower than the efficiency of a quantum Otto cycle
without parametric pumping. On the other hand, the quantum Otto cycle
has in turn a lower efficiency than a quantum Otto-type cycle with
fixed frequency. This explains why we choose to keep the QHO frequency
fixed such that the complete cycle only includes the Hamiltonian variation
when turning on and off the parametric pumping responsible for squeezing.
It is important to note that this procedure differs from the conventional
Otto cycle which consists of changing the frequency of the working
substance at the same time that a nonlinear term is turned on to modify
the Hamiltonian \citep{Karar2019,Mendes2021} .

Consider now the efficiency at maximum power in either
high or low temperature limits and considering the adiabatic condition
$Q^{*}=1$. For the high temperature
limit, after optimizing the net work in relation to the squeezing parameter we find $\Omega/\omega=\sqrt{\beta_{h}/\beta_{c}}$, from which
we obtain the well-known Ahlborn-Curzon relation \citep{Curzon1975, Esposito09maxpower, deffner2018efficiency,pena2023maxpower,Contreras-Vergara2023}.

\begin{equation}
\eta_{AC}=1-\sqrt{\frac{\beta_{h}}{\beta_{c}}}.
\end{equation}
For low temperatures $\beta_{c}\hbar\Omega/2\gg1$ and $\beta_{c}$
much greater than $\beta_{h}$, i.e., $\beta_{h}\rightarrow0$ we
obtain

\begin{equation}
\eta=1-\sqrt{\frac{\beta_{h}\hbar\Omega}{2}},\label{eq:16}
\end{equation}
which should be compared with Eq. (12) obtained in Ref.
\citep{Abah2012} that reads $\eta=1-\sqrt{\beta_{h}\hbar\omega_{1}/2}$,
$\omega_{1}$ being the lowest frequency of the QHO used as the working
substance of the Otto cycle. Interestingly, Eq. (\ref{eq:16}) shows
that given a minimum operating frequency $\omega_{1}$ at low temperatures,
we can always make the efficiency higher if, instead of changing the
frequency of the QHO, we couple this QHO with a parametric pumping,
since $\omega_{1}>\Omega$.

\begin{figure}
\centering{}\includegraphics[viewport=0bp 0bp 240bp 180bp]{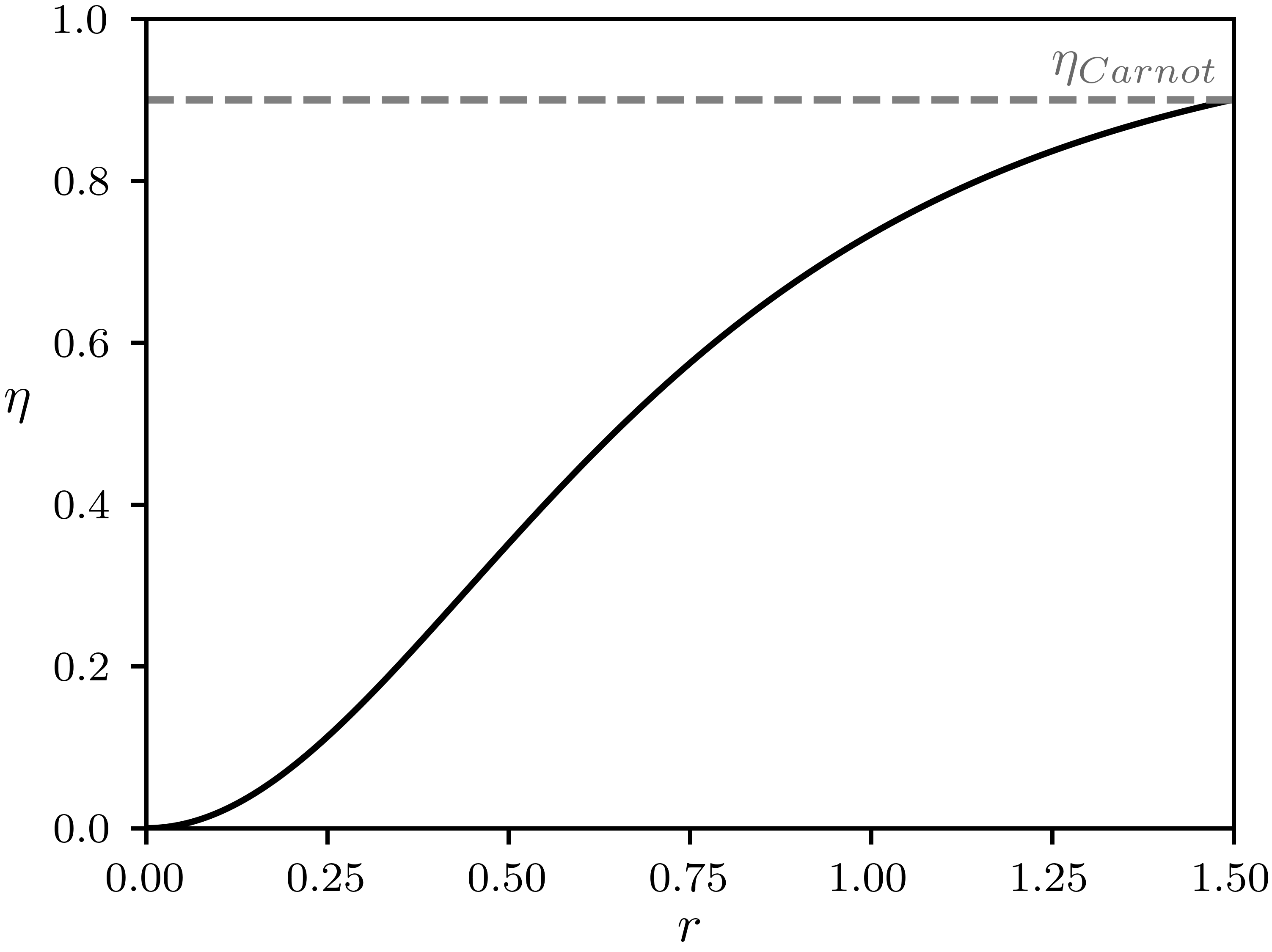}\caption{\label{fig:1}Otto engine efficiency versus $r$ for $Q^{*}=1$, $\omega=2\pi$,
$\beta_{c}=1$, and $\beta_{h}=0.1$. The Carnot limit given by the
dashed gray line is achieved when $r_{max}=\frac{1}{2}\cosh^{-1}\left(\beta_{c}/\beta_{h}\right)$.}
\end{figure}

Still considering the quasi-static process $Q^{*}=1$,
to further explore the effect of parametric pumping on the efficiency
of the Otto engine, in Fig. \ref{fig:1} we plot the efficiency versus
the squeezing parameter $r$ (solid blue line) for $\omega=2\pi$,
$\beta_{c}=1$, and $\beta_{h}=0.1$. The solid horizontal line refers
to the Carnot limit, which is achieved for $\chi\geq\pi$. This result
is in agreement with the work of  Refs. \citep{Karar2019} and
\citep{Mendes2021}, in which the authors show that anharmonicity
can improve the performance of quantum machines.

\subsection{\label{subsec:II.E}Efficiency at non-null power}

We now turn to the case where the Hamiltonian varies
non-slowly, i.e., where $Q^{*}\neq1$, and as a result the Otto-type
cycle runs at non-null power. For our purposes, we are going to perform a power analysis considering only the unitary strokes, since the thermalization strokes depend on the coupling between the system and the reservoir being considered. It is important to highlight, however, that for a real cycle, the state absorbs heat from the hot reservoir and dumps heat into the cold reservoir without complete thermalization occurring, which generally takes a long time compared to the time of the cycle.

First, let us consider a sudden change at high temperatures. For a sudden change in frequency, the adiabatic parameter is well-known $Q^* = (\omega^2 + \Omega^2)/2\omega\Omega$. For high temperatures we mean that $\beta_{\alpha}\hbar\omega\ll1$, $\alpha=c,h$. With this, the efficiency reads

\begin{equation}
\eta_{ss}=\frac{1-\sqrt{\beta_{h}/\beta_{c}}}{2+\sqrt{\beta_{h}/\beta_{c}}},
\end{equation}
which recovers Eq. (11) of Ref. \citep{Abah2012}. More interesting
is the analysis at low temperatures, where the quantum regime is evidenced.
For $\beta_{\alpha}\hbar\omega\gg1$, $\alpha=c,h$, we get

\begin{equation}
\eta_{ss}=\frac{1-\sqrt{\beta_{h}\hbar\Omega/2}}{2+\sqrt{\beta_{h}\hbar\Omega/2}},\label{eq:18}
\end{equation}
in contrast with Eq. (13) of Ref. \citep{Abah2012} where instead
of $\Omega$ they have the lowest frequency $\omega_{1}$. Again,
Eq. (\ref{eq:18}) shows that at low temperatures the efficiency of
an Otto cycle with fixed frequency but using a parametric pump is
always greater than the efficiency of a conventional quantum Otto
cycle.

It is not possible to overcome
Carnot by replacing the squeezing operation performed on the reservoir
with another squeezing operation performed directly on the working
substance. Furthermore, in the condition of non-adiabaticity $Q^* \neq1$, 
the efficiency never reaches the Carnot limit, an expected result and
different from what happens in the adiabatic case. That the Carnot
limit cannot be exceeded was shown in Ref. \citep{Campisi2014} making
use of fluctuation relations involving the non-equilibrium work and
heat exchanged with the reservoir. Taking advantage of the diagonalization
method applied here, in the next Section we will show this result
in a simplified alternative way.

\subsection{\label{subsec:2.F}Not beyond Carnot}

We will state the result in the form of the following
theorem:\medskip{}

\noindent {\textbf{Theorem}.}{\emph{
It is not possible to extract an efficiency $\eta$ greater than Carnot
efficiency $\eta_{carnot}$ from an Otto cycle without resorting to
quantum (non-thermal) reservoirs. }}{\medskip{}
}

To prove this theorem, we make use of the well-known
result according to which Hermitian operators can be diagonalized,
as for example was done previously doing a Bogoliubov transformation,
Eq. (\ref{eq:6}). In other words, it is always possible to write
\begin{equation}
H\left(a^{\dagger},a\right)=\hbar\Omega b^{\dagger}b,
\end{equation}
where $\Omega$ can of course have a different meaning from what we used before. Now, if we assume only thermal reservoirs and the Otto cycle, the
efficiency $\eta$ is given by Eq. \ref{eq:11} and it is always greater
when we keep the QHO frequency fixed. We therefore must show that
$\eta\le\eta_{Carnot}$. Taking into account Eq. (\ref{eq:9}) and
the engine condition $Q_{h}>0$, we can write 
\begin{equation}
\frac{\hbar\omega}{2}\left[\coth\left(\frac{\beta_{h}\hbar\omega}{2}\right)-Q^{*}\coth\left(\frac{\beta_{c}\hbar\Omega}{2}\right)\right]>0,
\end{equation}
from which we obtain
\begin{equation}
\frac{\coth\left(\frac{\beta_{h}\hbar\omega}{2}\right)}{\coth\left(\frac{\beta_{c}\hbar\Omega}{2}\right)}>Q^{*}\ge1,
\end{equation}
and therefore $\coth\left(\beta_{h}\omega/2\right)\ge\coth\left(\beta_{c}\Omega/2\right)$,
or, equivalently: 
\begin{equation}
\frac{\beta_{h}}{\beta_{c}}\leq\frac{\Omega}{\omega}.
\end{equation}
Now, multiplying both sides by the term in parentheses on the right
hand side by $\mathcal{F}$ of Eq. (\ref{eq:12})
\begin{equation}
\frac{\beta_{h}}{\beta_{c}}\mathcal{F}\leq\frac{\Omega}{\omega}\mathcal{F},
\end{equation}
and rearranging to
\begin{equation}
1-\frac{\beta_{h}}{\beta_{c}}\mathcal{F}\geq1-\frac{\Omega}{\omega}\mathcal{F},
\end{equation}
we can clearly see that the right hand side in the above inequality
is just $\eta$, as given by Eq. (\ref{eq:11}): 
\begin{equation}
1-\frac{\beta_{h}}{\beta_{c}}\left[\frac{\coth\left(\frac{\beta_{c}\hbar\Omega}{2}\right)-Q^{*}\coth\left(\frac{\beta_{h}\hbar\omega}{2}\right)}{Q^{*}\coth\left(\frac{\beta_{c}\hbar\Omega}{2}\right)-\coth\left(\frac{\beta_{h}\hbar\omega}{2}\right)}\right]\geq\eta.\label{eq:25}
\end{equation}

Consider now the following inequality:
\begin{equation}
\left(Q^{*}-1\right)\coth\left(\frac{\beta_{h}\hbar\omega}{2}\right)+\left(Q^{*}-1\right)\coth\left(\frac{\beta_{c}\hbar\Omega}{2}\right)\ge0,
\end{equation}
which is obvious since $Q^{*}\ge1$, and the $\coth$ function is
always positive for positive inputs. Then, from the above equation
we obtain 
\begin{equation}
\frac{Q^{*}\coth\left(\frac{\beta_{h}\hbar\omega}{2}\right)-\coth\left(\frac{\beta_{c}\hbar\Omega}{2}\right)}{\coth\left(\frac{\beta_{h}\hbar\omega}{2}\right)-Q^{*}\coth\left(\frac{\beta_{c}\hbar\Omega}{2}\right)}\ge1.
\end{equation}
Multiplying both sides by $\beta_{h}/\beta_{c}$ and rearranging:
\begin{equation}
1-\frac{\beta_{h}}{\beta_{c}}\left[\frac{Q^{*}\coth\left(\frac{\beta_{h}\hbar\omega}{2}\right)-\coth\left(\frac{\beta_{c}\hbar\Omega}{2}\right)}{\coth\left(\frac{\beta_{h}\hbar\omega}{2}\right)-Q^{*}\coth\left(\frac{\beta_{c}\hbar\Omega}{2}\right)}\right]\le1-\frac{\beta_{h}}{\beta_{c}}
\end{equation}
Note that the right hand side in the above equation is just $\eta_{carnot}$.
Finally, we use Eq. (\ref{eq:25}) to complete the proof: 
\begin{equation}
\eta\leq1-\frac{\beta_{h}}{\beta_{c}}\left[\frac{Q^{*}\coth\left(\frac{\beta_{h}\hbar\omega}{2}\right)-\coth\left(\frac{\beta_{c}\hbar\Omega}{2}\right)}{\coth\left(\frac{\beta_{h}\hbar\omega}{2}\right)-Q^{*}\coth\left(\frac{\beta_{c}\hbar\Omega}{2}\right)}\right]\le\eta_{Carnot}.
\end{equation}

\section{Numerical results}\label{sec:III}

\begin{figure*}
\begin{centering}
\includegraphics{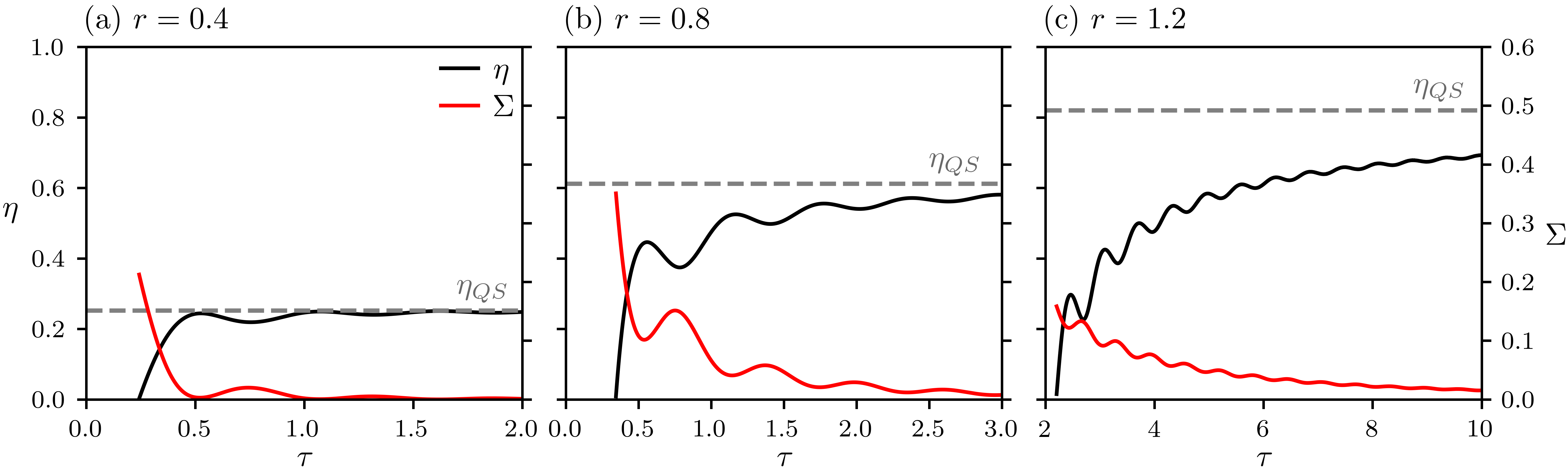}
\par\end{centering}
\caption{\label{fig:2}Efficiency $\eta=-W_{net}/Q_{abs}$ (black solid) and entropy production $\Sigma=D\left(\rho_{\tau}\parallel\rho_{i}\right)$ (red
solid) as a function of the time taken to perform the unitary strokes
for the squeezing parameters (a) $r=0.4$, (b) $r=0.8$, (c) $r=1.2$. 
In all figures $\omega=2\pi$, $\beta_{c}=1$ and $\beta_{h}=0.1$. The dashed horizontal curve is the efficiency $\eta_{QS}$ for the quasi-static case.}

\end{figure*}

We proceed to study the effect of non-adiabaticity in a
Otto-type cycle performed at non-null power without approximations
for temperatures. To this end, we use numerical methods in Python
and the QuTip toolbox \citep{Johansson2012,DelRe2012}. We run the
unitary expansion and compression strokes with the same linear function given by $\chi\left(t\right)=\chi\left(0\right)t$. In Figs. \ref{fig:2}(a)-(c)
we show how the efficiency behaves in terms of the duration of the
unitary strokes for squeezing parameters $r=0.4$, $r=0.8$, and $r=1.2$,
respectively.

\begin{figure}
\centering{}\includegraphics[viewport=0bp 0bp 240bp 180bp]{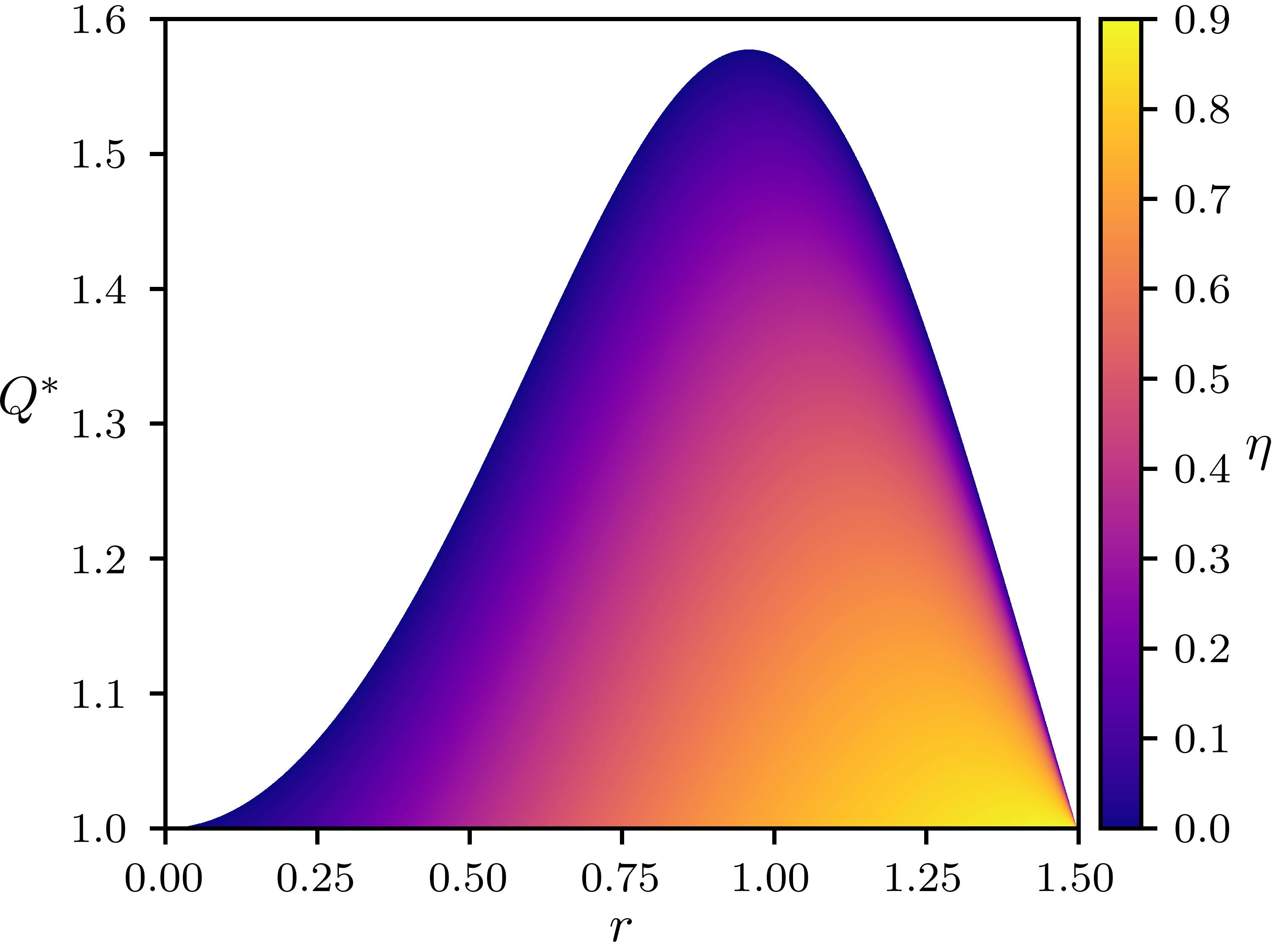}\caption{\label{fig:3}Efficiency as a function of both $\chi$ and $Q^{*}$.
Here we used $\omega=2\pi$, $\beta_{c}=1$ and $\beta_{h}=0.1$,
which makes Carnot efficiency $\eta=0.9$. Note that higher efficiencies
occur for $Q^{*}=1$ and $r>1.2$, in accordance with Fig. \ref{fig:1}.}
\end{figure}

Interestingly, notice how the efficiency oscillates
depending on the duration of the unitary stroke. Counter intuitively, this
means that depending on the times in which the unitary strokes are performed,
the efficiency can be higher even if the cycle is performed in a shorter
time as shown in Fig. \ref{fig:2}(a), and the quasi-static efficiency is achieved for some finite time cycles.  Also, note that the oscillation in efficiency values is greater
the greater the squeezing parameter. This non-trivial efficiency behavior
naturally leads us to ask about its causes. As is well known, one
of the causes for the decrease in useful work is the production of
entropy \citep{Deffner2010} due to the finite time of execution of
unitary strokes resulting in friction work \citep{Plastina2014} $\left\langle W_{fric}\right\rangle =-\frac{1}{\beta}_{i}D\left(\rho_{\tau}\parallel\rho_{i}\right)$.
Here, $\Sigma=D\left(\rho_{\tau}\parallel\rho_{i}\right)=tr\left\{ \rho_{\tau}\ln\rho_{\tau}\right\} -tr\left\{ \rho_{\tau}\ln\rho_{i}\right\} $
is the relative entropy, also known as the entropy production, $\rho_{\tau}$
is the density operator after the unitary stroke at a finite time
$\tau$ and $\rho_{i}$ is the thermalized density operator at inverse
temperature $\beta_{\alpha}$, $\alpha=c,h$. It would be expected,
therefore, that an increase (decrease) in entropy would lead to a
decrease (increase) in the useful work extracted from the engine,
which impacts efficiency, since it is directly proportional to the
net work extracted from the engine. Remarkably, from Figs.
\ref{fig:2}(a)-(c), we can see that the local maximums and minimums of the efficiency
coincide exactly with the local minima and maxima of entropy production.
As expected, the effect of performing the unitary strokes in finite
times is to decrease the efficiency of the engine operating at non-null
power. This non-null power is also captured by the adiabatic parameter
$Q^{*}$, whose increase, as we have seen, causes a decrease in engine
efficiency. To better visualize this, we plot in Fig. \ref{fig:3}
the efficiency $\eta$ as a function of both $r$ and $Q^{*}$.
Note that higher efficiencies occur for larger values of the squeezing
parameter and for values of $Q^{*}$ close to unity, corresponding
to the quasi-static case. Remarkably, we can note that even for
strokes performed at non-null power, $Q^{*}\neq1$, the efficiency
can be equal to or very close to Carnot's as long as the squeezing
parameter $r$ is large enough.

\section{Conclusion}

We investigate an Otto cycle powered by a squeezing operation whose
expansion and compression strokes are made with arbitrary speed. Applying
a Bogoliubov method to diagonalize the Hamiltonian we were able to
compare our results with others already known for the free quantum
harmonic oscillator (QHO). Unlike what was done previously, in which
the frequency of the QHO is varied, in our work we kept it fixed,
since this is the strategy that leads to better efficiency. Then,
we considered the effective frequency after diagonalization as the
new frequency of the system, which is dependent on the amplitude of
the parametric pump or, equivalently, on the squeezing parameter.
We then show that it is possible to achieve Carnot efficiency as long
as the squeezing parameter is large enough when considering the quasi-static case. Using our method, we were
able to show in an alternative way that the efficiency of the Otto
engine operating under two thermal reservoirs cannot overcome the
Carnot efficiency no matter the quantum resource introduced via Hamiltonian
of the system. Furthermore, we investigated entropy production and
showed that it perfectly explains the increase or decrease in efficiency
of the Otto-type engine introduced here.

\begin{acknowledgments}
We acknowledge financial support from the Brazilian agencies CAPES
(Financial code 001), CNPq, FAPEG, and São
Paulo Research Foundation (FAPESP), Grant No. 2021/04672-0. This work was performed as part
of the Brazilian National Institute of Science and Technology (INCT)
for Quantum Information\textbf{ }Grant No. 465469/2014-0.
\end{acknowledgments}

\bibliographystyle{apsrev4-2}
\bibliography{references}

\begin{thebibliography}{39}%
\makeatletter
\providecommand \@ifxundefined [1]{%
 \@ifx{#1\undefined}
}%
\providecommand \@ifnum [1]{%
 \ifnum #1\expandafter \@firstoftwo
 \else \expandafter \@secondoftwo
 \fi
}%
\providecommand \@ifx [1]{%
 \ifx #1\expandafter \@firstoftwo
 \else \expandafter \@secondoftwo
 \fi
}%
\providecommand \natexlab [1]{#1}%
\providecommand \enquote  [1]{``#1''}%
\providecommand \bibnamefont  [1]{#1}%
\providecommand \bibfnamefont [1]{#1}%
\providecommand \citenamefont [1]{#1}%
\providecommand \href@noop [0]{\@secondoftwo}%
\providecommand \href [0]{\begingroup \@sanitize@url \@href}%
\providecommand \@href[1]{\@@startlink{#1}\@@href}%
\providecommand \@@href[1]{\endgroup#1\@@endlink}%
\providecommand \@sanitize@url [0]{\catcode `\\12\catcode `\$12\catcode
  `\&12\catcode `\#12\catcode `\^12\catcode `\_12\catcode `\%12\relax}%
\providecommand \@@startlink[1]{}%
\providecommand \@@endlink[0]{}%
\providecommand \url  [0]{\begingroup\@sanitize@url \@url }%
\providecommand \@url [1]{\endgroup\@href {#1}{\urlprefix }}%
\providecommand \urlprefix  [0]{URL }%
\providecommand \Eprint [0]{\href }%
\providecommand \doibase [0]{https://doi.org/}%
\providecommand \selectlanguage [0]{\@gobble}%
\providecommand \bibinfo  [0]{\@secondoftwo}%
\providecommand \bibfield  [0]{\@secondoftwo}%
\providecommand \translation [1]{[#1]}%
\providecommand \BibitemOpen [0]{}%
\providecommand \bibitemStop [0]{}%
\providecommand \bibitemNoStop [0]{.\EOS\space}%
\providecommand \EOS [0]{\spacefactor3000\relax}%
\providecommand \BibitemShut  [1]{\csname bibitem#1\endcsname}%
\let\auto@bib@innerbib\@empty
\bibitem [{\citenamefont {Carnot}(1872)}]{Carnot1872}%
  \BibitemOpen
  \bibfield  {author} {\bibinfo {author} {\bibfnamefont {S.}~\bibnamefont
  {Carnot}},\ }in\ \href@noop {} {\emph {\bibinfo {booktitle} {Annales
  scientifiques de l'{\'E}cole Normale Sup{\'e}rieure}}},\ Vol.~\bibinfo
  {volume} {1}\ (\bibinfo {year} {1872})\ pp.\ \bibinfo {pages}
  {393--457}\BibitemShut {NoStop}%
\bibitem [{\citenamefont {Scovil}\ and\ \citenamefont
  {Schulz-DuBois}(1959)}]{Scovil1959}%
  \BibitemOpen
  \bibfield  {author} {\bibinfo {author} {\bibfnamefont {H.~E.}\ \bibnamefont
  {Scovil}}\ and\ \bibinfo {author} {\bibfnamefont {E.~O.}\ \bibnamefont
  {Schulz-DuBois}},\ }\href@noop {} {\bibfield  {journal} {\bibinfo  {journal}
  {Physical Review Letters}\ }\textbf {\bibinfo {volume} {2}},\ \bibinfo
  {pages} {262} (\bibinfo {year} {1959})}\BibitemShut {NoStop}%
\bibitem [{\citenamefont {Geusic}\ \emph {et~al.}(1967)\citenamefont {Geusic},
  \citenamefont {Schulz-DuBios},\ and\ \citenamefont {Scovil}}]{Geusic1967}%
  \BibitemOpen
  \bibfield  {author} {\bibinfo {author} {\bibfnamefont {J.}~\bibnamefont
  {Geusic}}, \bibinfo {author} {\bibfnamefont {E.}~\bibnamefont
  {Schulz-DuBios}},\ and\ \bibinfo {author} {\bibfnamefont {H.}~\bibnamefont
  {Scovil}},\ }\href@noop {} {\bibfield  {journal} {\bibinfo  {journal}
  {Physical Review}\ }\textbf {\bibinfo {volume} {156}},\ \bibinfo {pages}
  {343} (\bibinfo {year} {1967})}\BibitemShut {NoStop}%
\bibitem [{\citenamefont {Li}\ and\ \citenamefont {Jia}(2011)}]{Li2011}%
  \BibitemOpen
  \bibfield  {author} {\bibinfo {author} {\bibfnamefont {P.}~\bibnamefont
  {Li}}\ and\ \bibinfo {author} {\bibfnamefont {B.}~\bibnamefont {Jia}},\
  }\href@noop {} {\bibfield  {journal} {\bibinfo  {journal} {Physical Review
  E}\ }\textbf {\bibinfo {volume} {83}},\ \bibinfo {pages} {062104} (\bibinfo
  {year} {2011})}\BibitemShut {NoStop}%
\bibitem [{\citenamefont {Wright}\ \emph {et~al.}(2018)\citenamefont {Wright},
  \citenamefont {Gould}, \citenamefont {Carvalho}, \citenamefont {Bedkihal},\
  and\ \citenamefont {Vaccaro}}]{Wright2018}%
  \BibitemOpen
  \bibfield  {author} {\bibinfo {author} {\bibfnamefont {J.~S.}\ \bibnamefont
  {Wright}}, \bibinfo {author} {\bibfnamefont {T.}~\bibnamefont {Gould}},
  \bibinfo {author} {\bibfnamefont {A.~R.}\ \bibnamefont {Carvalho}}, \bibinfo
  {author} {\bibfnamefont {S.}~\bibnamefont {Bedkihal}},\ and\ \bibinfo
  {author} {\bibfnamefont {J.~A.}\ \bibnamefont {Vaccaro}},\ }\href@noop {}
  {\bibfield  {journal} {\bibinfo  {journal} {Physical Review A}\ }\textbf
  {\bibinfo {volume} {97}},\ \bibinfo {pages} {052104} (\bibinfo {year}
  {2018})}\BibitemShut {NoStop}%
\bibitem [{\citenamefont {N{\"u}{\ss}eler}\ \emph {et~al.}(2020)\citenamefont
  {N{\"u}{\ss}eler}, \citenamefont {Dhand}, \citenamefont {Huelga},\ and\
  \citenamefont {Plenio}}]{NuBeler2020}%
  \BibitemOpen
  \bibfield  {author} {\bibinfo {author} {\bibfnamefont {A.}~\bibnamefont
  {N{\"u}{\ss}eler}}, \bibinfo {author} {\bibfnamefont {I.}~\bibnamefont
  {Dhand}}, \bibinfo {author} {\bibfnamefont {S.~F.}\ \bibnamefont {Huelga}},\
  and\ \bibinfo {author} {\bibfnamefont {M.~B.}\ \bibnamefont {Plenio}},\
  }\href@noop {} {\bibfield  {journal} {\bibinfo  {journal} {Physical Review
  B}\ }\textbf {\bibinfo {volume} {101}},\ \bibinfo {pages} {155134} (\bibinfo
  {year} {2020})}\BibitemShut {NoStop}%
\bibitem [{\citenamefont {Camati}\ \emph {et~al.}(2019)\citenamefont {Camati},
  \citenamefont {Santos},\ and\ \citenamefont {Serra}}]{Camati2019}%
  \BibitemOpen
  \bibfield  {author} {\bibinfo {author} {\bibfnamefont {P.~A.}\ \bibnamefont
  {Camati}}, \bibinfo {author} {\bibfnamefont {J.~F.}\ \bibnamefont {Santos}},\
  and\ \bibinfo {author} {\bibfnamefont {R.~M.}\ \bibnamefont {Serra}},\
  }\href@noop {} {\bibfield  {journal} {\bibinfo  {journal} {Physical Review
  A}\ }\textbf {\bibinfo {volume} {99}},\ \bibinfo {pages} {062103} (\bibinfo
  {year} {2019})}\BibitemShut {NoStop}%
\bibitem [{\citenamefont {Herrera}\ \emph {et~al.}(2022)\citenamefont
  {Herrera}, \citenamefont {Reina}, \citenamefont {D'Amico},\ and\
  \citenamefont {Serra}}]{Herrera2022}%
  \BibitemOpen
  \bibfield  {author} {\bibinfo {author} {\bibfnamefont {M.}~\bibnamefont
  {Herrera}}, \bibinfo {author} {\bibfnamefont {J.~H.}\ \bibnamefont {Reina}},
  \bibinfo {author} {\bibfnamefont {I.}~\bibnamefont {D'Amico}},\ and\ \bibinfo
  {author} {\bibfnamefont {R.~M.}\ \bibnamefont {Serra}},\ }\href@noop {}
  {\bibfield  {journal} {\bibinfo  {journal} {arXiv preprint arXiv:2211.11449}\
  } (\bibinfo {year} {2022})}\BibitemShut {NoStop}%
\bibitem [{\citenamefont {Ro{\ss}nagel}\ \emph {et~al.}(2014)\citenamefont
  {Ro{\ss}nagel}, \citenamefont {Abah}, \citenamefont {Schmidt-Kaler},
  \citenamefont {Singer},\ and\ \citenamefont {Lutz}}]{RoBnagel2014}%
  \BibitemOpen
  \bibfield  {author} {\bibinfo {author} {\bibfnamefont {J.}~\bibnamefont
  {Ro{\ss}nagel}}, \bibinfo {author} {\bibfnamefont {O.}~\bibnamefont {Abah}},
  \bibinfo {author} {\bibfnamefont {F.}~\bibnamefont {Schmidt-Kaler}}, \bibinfo
  {author} {\bibfnamefont {K.}~\bibnamefont {Singer}},\ and\ \bibinfo {author}
  {\bibfnamefont {E.}~\bibnamefont {Lutz}},\ }\href@noop {} {\bibfield
  {journal} {\bibinfo  {journal} {Physical review letters}\ }\textbf {\bibinfo
  {volume} {112}},\ \bibinfo {pages} {030602} (\bibinfo {year}
  {2014})}\BibitemShut {NoStop}%
\bibitem [{\citenamefont {de~Assis}\ \emph {et~al.}(2021)\citenamefont
  {de~Assis}, \citenamefont {Sales}, \citenamefont {Mendes},\ and\
  \citenamefont {de~Almeida}}]{Assis2021}%
  \BibitemOpen
  \bibfield  {author} {\bibinfo {author} {\bibfnamefont {R.~J.}\ \bibnamefont
  {de~Assis}}, \bibinfo {author} {\bibfnamefont {J.~S.}\ \bibnamefont {Sales}},
  \bibinfo {author} {\bibfnamefont {U.~C.}\ \bibnamefont {Mendes}},\ and\
  \bibinfo {author} {\bibfnamefont {N.~G.}\ \bibnamefont {de~Almeida}},\
  }\href@noop {} {\bibfield  {journal} {\bibinfo  {journal} {Journal of Physics
  B: Atomic, Molecular and Optical Physics}\ }\textbf {\bibinfo {volume}
  {54}},\ \bibinfo {pages} {095501} (\bibinfo {year} {2021})}\BibitemShut
  {NoStop}%
\bibitem [{\citenamefont {Long}\ and\ \citenamefont {Liu}(2015)}]{Long2015}%
  \BibitemOpen
  \bibfield  {author} {\bibinfo {author} {\bibfnamefont {R.}~\bibnamefont
  {Long}}\ and\ \bibinfo {author} {\bibfnamefont {W.}~\bibnamefont {Liu}},\
  }\href@noop {} {\bibfield  {journal} {\bibinfo  {journal} {Physical Review
  E}\ }\textbf {\bibinfo {volume} {91}},\ \bibinfo {pages} {062137} (\bibinfo
  {year} {2015})}\BibitemShut {NoStop}%
\bibitem [{\citenamefont {Klaers}\ \emph {et~al.}(2017)\citenamefont {Klaers},
  \citenamefont {Faelt}, \citenamefont {Imamoglu},\ and\ \citenamefont
  {Togan}}]{Klaers2017}%
  \BibitemOpen
  \bibfield  {author} {\bibinfo {author} {\bibfnamefont {J.}~\bibnamefont
  {Klaers}}, \bibinfo {author} {\bibfnamefont {S.}~\bibnamefont {Faelt}},
  \bibinfo {author} {\bibfnamefont {A.}~\bibnamefont {Imamoglu}},\ and\
  \bibinfo {author} {\bibfnamefont {E.}~\bibnamefont {Togan}},\ }\href@noop {}
  {\bibfield  {journal} {\bibinfo  {journal} {Physical Review X}\ }\textbf
  {\bibinfo {volume} {7}},\ \bibinfo {pages} {031044} (\bibinfo {year}
  {2017})}\BibitemShut {NoStop}%
\bibitem [{\citenamefont {Wang}\ \emph {et~al.}(2019)\citenamefont {Wang},
  \citenamefont {He},\ and\ \citenamefont {Ma}}]{Wang2019}%
  \BibitemOpen
  \bibfield  {author} {\bibinfo {author} {\bibfnamefont {J.}~\bibnamefont
  {Wang}}, \bibinfo {author} {\bibfnamefont {J.}~\bibnamefont {He}},\ and\
  \bibinfo {author} {\bibfnamefont {Y.}~\bibnamefont {Ma}},\ }\href@noop {}
  {\bibfield  {journal} {\bibinfo  {journal} {Physical Review E}\ }\textbf
  {\bibinfo {volume} {100}},\ \bibinfo {pages} {052126} (\bibinfo {year}
  {2019})}\BibitemShut {NoStop}%
\bibitem [{\citenamefont {de~Assis}\ \emph {et~al.}(2019)\citenamefont
  {de~Assis}, \citenamefont {de~Mendon{\c{c}}a}, \citenamefont {Villas-Boas},
  \citenamefont {de~Souza}, \citenamefont {Sarthour}, \citenamefont
  {Oliveira},\ and\ \citenamefont {de~Almeida}}]{Assis2019}%
  \BibitemOpen
  \bibfield  {author} {\bibinfo {author} {\bibfnamefont {R.~J.}\ \bibnamefont
  {de~Assis}}, \bibinfo {author} {\bibfnamefont {T.~M.}\ \bibnamefont
  {de~Mendon{\c{c}}a}}, \bibinfo {author} {\bibfnamefont {C.~J.}\ \bibnamefont
  {Villas-Boas}}, \bibinfo {author} {\bibfnamefont {A.~M.}\ \bibnamefont
  {de~Souza}}, \bibinfo {author} {\bibfnamefont {R.~S.}\ \bibnamefont
  {Sarthour}}, \bibinfo {author} {\bibfnamefont {I.~S.}\ \bibnamefont
  {Oliveira}},\ and\ \bibinfo {author} {\bibfnamefont {N.~G.}\ \bibnamefont
  {de~Almeida}},\ }\href@noop {} {\bibfield  {journal} {\bibinfo  {journal}
  {Physical Review Letters}\ }\textbf {\bibinfo {volume} {122}},\ \bibinfo
  {pages} {240602} (\bibinfo {year} {2019})}\BibitemShut {NoStop}%
\bibitem [{\citenamefont {Lee}\ \emph {et~al.}(2020)\citenamefont {Lee},
  \citenamefont {Ha}, \citenamefont {Park},\ and\ \citenamefont
  {Jeong}}]{Lee2020}%
  \BibitemOpen
  \bibfield  {author} {\bibinfo {author} {\bibfnamefont {S.}~\bibnamefont
  {Lee}}, \bibinfo {author} {\bibfnamefont {M.}~\bibnamefont {Ha}}, \bibinfo
  {author} {\bibfnamefont {J.-M.}\ \bibnamefont {Park}},\ and\ \bibinfo
  {author} {\bibfnamefont {H.}~\bibnamefont {Jeong}},\ }\href@noop {}
  {\bibfield  {journal} {\bibinfo  {journal} {Physical Review E}\ }\textbf
  {\bibinfo {volume} {101}},\ \bibinfo {pages} {022127} (\bibinfo {year}
  {2020})}\BibitemShut {NoStop}%
\bibitem [{\citenamefont {Damas}\ \emph {et~al.}(2022)\citenamefont {Damas},
  \citenamefont {de~Assis},\ and\ \citenamefont {de~Almeida}}]{Damas2022}%
  \BibitemOpen
  \bibfield  {author} {\bibinfo {author} {\bibfnamefont {G.~G.}\ \bibnamefont
  {Damas}}, \bibinfo {author} {\bibfnamefont {R.~J.}\ \bibnamefont
  {de~Assis}},\ and\ \bibinfo {author} {\bibfnamefont {N.~G.}\ \bibnamefont
  {de~Almeida}},\ }\href@noop {} {\bibfield  {journal} {\bibinfo  {journal}
  {arXiv preprint arXiv:2207.08862}\ } (\bibinfo {year} {2022})}\BibitemShut
  {NoStop}%
\bibitem [{\citenamefont {Nettersheim}\ \emph {et~al.}(2022)\citenamefont
  {Nettersheim}, \citenamefont {Burgardt}, \citenamefont {Bouton},
  \citenamefont {Adam}, \citenamefont {Lutz},\ and\ \citenamefont
  {Widera}}]{Nettersheim2022}%
  \BibitemOpen
  \bibfield  {author} {\bibinfo {author} {\bibfnamefont {J.}~\bibnamefont
  {Nettersheim}}, \bibinfo {author} {\bibfnamefont {S.}~\bibnamefont
  {Burgardt}}, \bibinfo {author} {\bibfnamefont {Q.}~\bibnamefont {Bouton}},
  \bibinfo {author} {\bibfnamefont {D.}~\bibnamefont {Adam}}, \bibinfo {author}
  {\bibfnamefont {E.}~\bibnamefont {Lutz}},\ and\ \bibinfo {author}
  {\bibfnamefont {A.}~\bibnamefont {Widera}},\ }\href@noop {} {\bibfield
  {journal} {\bibinfo  {journal} {arXiv preprint arXiv:2207.09272}\ } (\bibinfo
  {year} {2022})}\BibitemShut {NoStop}%
\bibitem [{\citenamefont {Campisi}(2014)}]{Campisi2014}%
  \BibitemOpen
  \bibfield  {author} {\bibinfo {author} {\bibfnamefont {M.}~\bibnamefont
  {Campisi}},\ }\href@noop {} {\bibfield  {journal} {\bibinfo  {journal}
  {Journal of Physics A: Mathematical and Theoretical}\ }\textbf {\bibinfo
  {volume} {47}},\ \bibinfo {pages} {245001} (\bibinfo {year}
  {2014})}\BibitemShut {NoStop}%
\bibitem [{\citenamefont {De~Almeida}\ \emph {et~al.}(2004)\citenamefont
  {De~Almeida}, \citenamefont {Serra}, \citenamefont {Villas-Boas},\ and\
  \citenamefont {Moussa}}]{Almeida2004}%
  \BibitemOpen
  \bibfield  {author} {\bibinfo {author} {\bibfnamefont {N.}~\bibnamefont
  {De~Almeida}}, \bibinfo {author} {\bibfnamefont {R.}~\bibnamefont {Serra}},
  \bibinfo {author} {\bibfnamefont {C.}~\bibnamefont {Villas-Boas}},\ and\
  \bibinfo {author} {\bibfnamefont {M.}~\bibnamefont {Moussa}},\ }\href@noop {}
  {\bibfield  {journal} {\bibinfo  {journal} {Physical Review A}\ }\textbf
  {\bibinfo {volume} {69}},\ \bibinfo {pages} {035802} (\bibinfo {year}
  {2004})}\BibitemShut {NoStop}%
\bibitem [{\citenamefont {Villas-Boas}\ \emph {et~al.}(2003)\citenamefont
  {Villas-Boas}, \citenamefont {de~Almeida}, \citenamefont {Serra},\ and\
  \citenamefont {Moussa}}]{Villas-Boas2003}%
  \BibitemOpen
  \bibfield  {author} {\bibinfo {author} {\bibfnamefont {C.}~\bibnamefont
  {Villas-Boas}}, \bibinfo {author} {\bibfnamefont {N.}~\bibnamefont
  {de~Almeida}}, \bibinfo {author} {\bibfnamefont {R.}~\bibnamefont {Serra}},\
  and\ \bibinfo {author} {\bibfnamefont {M.}~\bibnamefont {Moussa}},\
  }\href@noop {} {\bibfield  {journal} {\bibinfo  {journal} {Physical Review
  A}\ }\textbf {\bibinfo {volume} {68}},\ \bibinfo {pages} {061801} (\bibinfo
  {year} {2003})}\BibitemShut {NoStop}%
\bibitem [{\citenamefont {Vogel}\ and\ \citenamefont
  {de~Matos~Filho}(1995)}]{Vogel1995}%
  \BibitemOpen
  \bibfield  {author} {\bibinfo {author} {\bibfnamefont {W.}~\bibnamefont
  {Vogel}}\ and\ \bibinfo {author} {\bibfnamefont {R.}~\bibnamefont
  {de~Matos~Filho}},\ }\href@noop {} {\bibfield  {journal} {\bibinfo  {journal}
  {Physical Review A}\ }\textbf {\bibinfo {volume} {52}},\ \bibinfo {pages}
  {4214} (\bibinfo {year} {1995})}\BibitemShut {NoStop}%
\bibitem [{\citenamefont {Blais}\ \emph {et~al.}(2021)\citenamefont {Blais},
  \citenamefont {Grimsmo}, \citenamefont {Girvin},\ and\ \citenamefont
  {Wallraff}}]{Blais2020}%
  \BibitemOpen
  \bibfield  {author} {\bibinfo {author} {\bibfnamefont {A.}~\bibnamefont
  {Blais}}, \bibinfo {author} {\bibfnamefont {A.~L.}\ \bibnamefont {Grimsmo}},
  \bibinfo {author} {\bibfnamefont {S.~M.}\ \bibnamefont {Girvin}},\ and\
  \bibinfo {author} {\bibfnamefont {A.}~\bibnamefont {Wallraff}},\ }\href@noop
  {} {\bibfield  {journal} {\bibinfo  {journal} {Reviews of Modern Physics}\
  }\textbf {\bibinfo {volume} {93}},\ \bibinfo {pages} {025005} (\bibinfo
  {year} {2021})}\BibitemShut {NoStop}%
\bibitem [{\citenamefont {Miwa}\ \emph {et~al.}(2014)\citenamefont {Miwa},
  \citenamefont {Yoshikawa}, \citenamefont {Iwata}, \citenamefont {Endo},
  \citenamefont {Marek}, \citenamefont {Filip}, \citenamefont {van Loock},\
  and\ \citenamefont {Furusawa}}]{miwa2014expsqueez}%
  \BibitemOpen
  \bibfield  {author} {\bibinfo {author} {\bibfnamefont {Y.}~\bibnamefont
  {Miwa}}, \bibinfo {author} {\bibfnamefont {J.-i.}\ \bibnamefont {Yoshikawa}},
  \bibinfo {author} {\bibfnamefont {N.}~\bibnamefont {Iwata}}, \bibinfo
  {author} {\bibfnamefont {M.}~\bibnamefont {Endo}}, \bibinfo {author}
  {\bibfnamefont {P.}~\bibnamefont {Marek}}, \bibinfo {author} {\bibfnamefont
  {R.}~\bibnamefont {Filip}}, \bibinfo {author} {\bibfnamefont
  {P.}~\bibnamefont {van Loock}},\ and\ \bibinfo {author} {\bibfnamefont
  {A.}~\bibnamefont {Furusawa}},\ }\href@noop {} {\bibfield  {journal}
  {\bibinfo  {journal} {Physical Review Letters}\ }\textbf {\bibinfo {volume}
  {113}},\ \bibinfo {pages} {013601} (\bibinfo {year} {2014})}\BibitemShut
  {NoStop}%
\bibitem [{\citenamefont {Yoshikawa}\ \emph {et~al.}(2007)\citenamefont
  {Yoshikawa}, \citenamefont {Hayashi}, \citenamefont {Akiyama}, \citenamefont
  {Takei}, \citenamefont {Huck}, \citenamefont {Andersen},\ and\ \citenamefont
  {Furusawa}}]{Yoshikawa07expsqueez}%
  \BibitemOpen
  \bibfield  {author} {\bibinfo {author} {\bibfnamefont {J.-i.}\ \bibnamefont
  {Yoshikawa}}, \bibinfo {author} {\bibfnamefont {T.}~\bibnamefont {Hayashi}},
  \bibinfo {author} {\bibfnamefont {T.}~\bibnamefont {Akiyama}}, \bibinfo
  {author} {\bibfnamefont {N.}~\bibnamefont {Takei}}, \bibinfo {author}
  {\bibfnamefont {A.}~\bibnamefont {Huck}}, \bibinfo {author} {\bibfnamefont
  {U.~L.}\ \bibnamefont {Andersen}},\ and\ \bibinfo {author} {\bibfnamefont
  {A.}~\bibnamefont {Furusawa}},\ }\href
  {https://doi.org/10.1103/PhysRevA.76.060301} {\bibfield  {journal} {\bibinfo
  {journal} {Phys. Rev. A}\ }\textbf {\bibinfo {volume} {76}},\ \bibinfo
  {pages} {060301} (\bibinfo {year} {2007})}\BibitemShut {NoStop}%
\bibitem [{\citenamefont {Miyata}\ \emph {et~al.}(2014)\citenamefont {Miyata},
  \citenamefont {Ogawa}, \citenamefont {Marek}, \citenamefont {Filip},
  \citenamefont {Yonezawa}, \citenamefont {Yoshikawa},\ and\ \citenamefont
  {Furusawa}}]{miyata2014expsqueez}%
  \BibitemOpen
  \bibfield  {author} {\bibinfo {author} {\bibfnamefont {K.}~\bibnamefont
  {Miyata}}, \bibinfo {author} {\bibfnamefont {H.}~\bibnamefont {Ogawa}},
  \bibinfo {author} {\bibfnamefont {P.}~\bibnamefont {Marek}}, \bibinfo
  {author} {\bibfnamefont {R.}~\bibnamefont {Filip}}, \bibinfo {author}
  {\bibfnamefont {H.}~\bibnamefont {Yonezawa}}, \bibinfo {author}
  {\bibfnamefont {J.-i.}\ \bibnamefont {Yoshikawa}},\ and\ \bibinfo {author}
  {\bibfnamefont {A.}~\bibnamefont {Furusawa}},\ }\href@noop {} {\bibfield
  {journal} {\bibinfo  {journal} {Physical Review A}\ }\textbf {\bibinfo
  {volume} {90}},\ \bibinfo {pages} {060302} (\bibinfo {year}
  {2014})}\BibitemShut {NoStop}%
\bibitem [{\citenamefont {Abah}\ \emph {et~al.}(2012)\citenamefont {Abah},
  \citenamefont {Rossnagel}, \citenamefont {Jacob}, \citenamefont {Deffner},
  \citenamefont {Schmidt-Kaler}, \citenamefont {Singer},\ and\ \citenamefont
  {Lutz}}]{Abah2012}%
  \BibitemOpen
  \bibfield  {author} {\bibinfo {author} {\bibfnamefont {O.}~\bibnamefont
  {Abah}}, \bibinfo {author} {\bibfnamefont {J.}~\bibnamefont {Rossnagel}},
  \bibinfo {author} {\bibfnamefont {G.}~\bibnamefont {Jacob}}, \bibinfo
  {author} {\bibfnamefont {S.}~\bibnamefont {Deffner}}, \bibinfo {author}
  {\bibfnamefont {F.}~\bibnamefont {Schmidt-Kaler}}, \bibinfo {author}
  {\bibfnamefont {K.}~\bibnamefont {Singer}},\ and\ \bibinfo {author}
  {\bibfnamefont {E.}~\bibnamefont {Lutz}},\ }\href@noop {} {\bibfield
  {journal} {\bibinfo  {journal} {Physical review letters}\ }\textbf {\bibinfo
  {volume} {109}},\ \bibinfo {pages} {203006} (\bibinfo {year}
  {2012})}\BibitemShut {NoStop}%
\bibitem [{\citenamefont {Alicki}(1979)}]{Alicki1979}%
  \BibitemOpen
  \bibfield  {author} {\bibinfo {author} {\bibfnamefont {R.}~\bibnamefont
  {Alicki}},\ }\href@noop {} {\bibfield  {journal} {\bibinfo  {journal}
  {Journal of Physics A: Mathematical and General}\ }\textbf {\bibinfo {volume}
  {12}},\ \bibinfo {pages} {L103} (\bibinfo {year} {1979})}\BibitemShut
  {NoStop}%
\bibitem [{\citenamefont {Husimi}(1953)}]{Husimi1953}%
  \BibitemOpen
  \bibfield  {author} {\bibinfo {author} {\bibfnamefont {K.}~\bibnamefont
  {Husimi}},\ }\href@noop {} {\bibfield  {journal} {\bibinfo  {journal}
  {Progress of Theoretical Physics}\ }\textbf {\bibinfo {volume} {9}},\
  \bibinfo {pages} {381} (\bibinfo {year} {1953})}\BibitemShut {NoStop}%
\bibitem [{\citenamefont {Karar}\ \emph {et~al.}(2019)\citenamefont {Karar},
  \citenamefont {Datta}, \citenamefont {Ghosh},\ and\ \citenamefont
  {Majumdar}}]{Karar2019}%
  \BibitemOpen
  \bibfield  {author} {\bibinfo {author} {\bibfnamefont {S.}~\bibnamefont
  {Karar}}, \bibinfo {author} {\bibfnamefont {S.}~\bibnamefont {Datta}},
  \bibinfo {author} {\bibfnamefont {S.}~\bibnamefont {Ghosh}},\ and\ \bibinfo
  {author} {\bibfnamefont {A.}~\bibnamefont {Majumdar}},\ }\href@noop {}
  {\bibfield  {journal} {\bibinfo  {journal} {arXiv preprint arXiv:1902.10616}\
  } (\bibinfo {year} {2019})}\BibitemShut {NoStop}%
\bibitem [{\citenamefont {Mendes}\ \emph {et~al.}(2021)\citenamefont {Mendes},
  \citenamefont {Sales},\ and\ \citenamefont {de~Almeida}}]{Mendes2021}%
  \BibitemOpen
  \bibfield  {author} {\bibinfo {author} {\bibfnamefont {U.~C.}\ \bibnamefont
  {Mendes}}, \bibinfo {author} {\bibfnamefont {J.~S.}\ \bibnamefont {Sales}},\
  and\ \bibinfo {author} {\bibfnamefont {N.~G.}\ \bibnamefont {de~Almeida}},\
  }\href@noop {} {\bibfield  {journal} {\bibinfo  {journal} {Journal of Physics
  B: Atomic, Molecular and Optical Physics}\ }\textbf {\bibinfo {volume}
  {54}},\ \bibinfo {pages} {175504} (\bibinfo {year} {2021})}\BibitemShut
  {NoStop}%
\bibitem [{\citenamefont {Curzon}\ and\ \citenamefont
  {Ahlborn}(1975)}]{Curzon1975}%
  \BibitemOpen
  \bibfield  {author} {\bibinfo {author} {\bibfnamefont {F.~L.}\ \bibnamefont
  {Curzon}}\ and\ \bibinfo {author} {\bibfnamefont {B.}~\bibnamefont
  {Ahlborn}},\ }\href@noop {} {\bibfield  {journal} {\bibinfo  {journal}
  {American Journal of Physics}\ }\textbf {\bibinfo {volume} {43}},\ \bibinfo
  {pages} {22} (\bibinfo {year} {1975})}\BibitemShut {NoStop}%
\bibitem [{\citenamefont {Esposito}\ \emph {et~al.}(2009)\citenamefont
  {Esposito}, \citenamefont {Lindenberg},\ and\ \citenamefont {Van~den
  Broeck}}]{Esposito09maxpower}%
  \BibitemOpen
  \bibfield  {author} {\bibinfo {author} {\bibfnamefont {M.}~\bibnamefont
  {Esposito}}, \bibinfo {author} {\bibfnamefont {K.}~\bibnamefont
  {Lindenberg}},\ and\ \bibinfo {author} {\bibfnamefont {C.}~\bibnamefont
  {Van~den Broeck}},\ }\href {https://doi.org/10.1103/PhysRevLett.102.130602}
  {\bibfield  {journal} {\bibinfo  {journal} {Phys. Rev. Lett.}\ }\textbf
  {\bibinfo {volume} {102}},\ \bibinfo {pages} {130602} (\bibinfo {year}
  {2009})}\BibitemShut {NoStop}%
\bibitem [{\citenamefont {Deffner}(2018)}]{deffner2018efficiency}%
  \BibitemOpen
  \bibfield  {author} {\bibinfo {author} {\bibfnamefont {S.}~\bibnamefont
  {Deffner}},\ }\href@noop {} {\bibfield  {journal} {\bibinfo  {journal}
  {Entropy}\ }\textbf {\bibinfo {volume} {20}},\ \bibinfo {pages} {875}
  (\bibinfo {year} {2018})}\BibitemShut {NoStop}%
\bibitem [{\citenamefont {Pe{\~n}a}\ \emph {et~al.}(2023)\citenamefont
  {Pe{\~n}a}, \citenamefont {Myers}, \citenamefont {{\'O}rdenes}, \citenamefont
  {Albarr{\'a}n-Arriagada},\ and\ \citenamefont {Vargas}}]{pena2023maxpower}%
  \BibitemOpen
  \bibfield  {author} {\bibinfo {author} {\bibfnamefont {F.~J.}\ \bibnamefont
  {Pe{\~n}a}}, \bibinfo {author} {\bibfnamefont {N.~M.}\ \bibnamefont {Myers}},
  \bibinfo {author} {\bibfnamefont {D.}~\bibnamefont {{\'O}rdenes}}, \bibinfo
  {author} {\bibfnamefont {F.}~\bibnamefont {Albarr{\'a}n-Arriagada}},\ and\
  \bibinfo {author} {\bibfnamefont {P.}~\bibnamefont {Vargas}},\ }\href@noop {}
  {\bibfield  {journal} {\bibinfo  {journal} {Entropy}\ }\textbf {\bibinfo
  {volume} {25}},\ \bibinfo {pages} {518} (\bibinfo {year} {2023})}\BibitemShut
  {NoStop}%
\bibitem [{\citenamefont {Contreras-Vergara}\ \emph {et~al.}(2023)\citenamefont
  {Contreras-Vergara}, \citenamefont {S\'anchez-Salas}, \citenamefont
  {Valencia-Ortega},\ and\ \citenamefont
  {Jim\'enez-Aquino}}]{Contreras-Vergara2023}%
  \BibitemOpen
  \bibfield  {author} {\bibinfo {author} {\bibfnamefont {O.}~\bibnamefont
  {Contreras-Vergara}}, \bibinfo {author} {\bibfnamefont {N.}~\bibnamefont
  {S\'anchez-Salas}}, \bibinfo {author} {\bibfnamefont {G.}~\bibnamefont
  {Valencia-Ortega}},\ and\ \bibinfo {author} {\bibfnamefont {J.~I.}\
  \bibnamefont {Jim\'enez-Aquino}},\ }\href
  {https://doi.org/10.1103/PhysRevE.108.014123} {\bibfield  {journal} {\bibinfo
   {journal} {Phys. Rev. E}\ }\textbf {\bibinfo {volume} {108}},\ \bibinfo
  {pages} {014123} (\bibinfo {year} {2023})}\BibitemShut {NoStop}%
\bibitem [{\citenamefont {Johansson}\ \emph {et~al.}(2012)\citenamefont
  {Johansson}, \citenamefont {Nation},\ and\ \citenamefont
  {Nori}}]{Johansson2012}%
  \BibitemOpen
  \bibfield  {author} {\bibinfo {author} {\bibfnamefont {J.~R.}\ \bibnamefont
  {Johansson}}, \bibinfo {author} {\bibfnamefont {P.~D.}\ \bibnamefont
  {Nation}},\ and\ \bibinfo {author} {\bibfnamefont {F.}~\bibnamefont {Nori}},\
  }\href@noop {} {\bibfield  {journal} {\bibinfo  {journal} {Computer Physics
  Communications}\ }\textbf {\bibinfo {volume} {183}},\ \bibinfo {pages} {1760}
  (\bibinfo {year} {2012})}\BibitemShut {NoStop}%
\bibitem [{\citenamefont {Del~Re}\ \emph {et~al.}(2020)\citenamefont {Del~Re},
  \citenamefont {Rost}, \citenamefont {Kemper},\ and\ \citenamefont
  {Freericks}}]{DelRe2012}%
  \BibitemOpen
  \bibfield  {author} {\bibinfo {author} {\bibfnamefont {L.}~\bibnamefont
  {Del~Re}}, \bibinfo {author} {\bibfnamefont {B.}~\bibnamefont {Rost}},
  \bibinfo {author} {\bibfnamefont {A.}~\bibnamefont {Kemper}},\ and\ \bibinfo
  {author} {\bibfnamefont {J.}~\bibnamefont {Freericks}},\ }\href@noop {}
  {\bibfield  {journal} {\bibinfo  {journal} {Physical Review B}\ }\textbf
  {\bibinfo {volume} {102}},\ \bibinfo {pages} {125112} (\bibinfo {year}
  {2020})}\BibitemShut {NoStop}%
\bibitem [{\citenamefont {Deffner}\ and\ \citenamefont
  {Lutz}(2010)}]{Deffner2010}%
  \BibitemOpen
  \bibfield  {author} {\bibinfo {author} {\bibfnamefont {S.}~\bibnamefont
  {Deffner}}\ and\ \bibinfo {author} {\bibfnamefont {E.}~\bibnamefont {Lutz}},\
  }\href {https://doi.org/10.1103/PhysRevLett.105.170402} {\bibfield  {journal}
  {\bibinfo  {journal} {Phys. Rev. Lett.}\ }\textbf {\bibinfo {volume} {105}},\
  \bibinfo {pages} {170402} (\bibinfo {year} {2010})}\BibitemShut {NoStop}%
\bibitem [{\citenamefont {Plastina}\ \emph {et~al.}(2014)\citenamefont
  {Plastina}, \citenamefont {Alecce}, \citenamefont {Apollaro}, \citenamefont
  {Falcone}, \citenamefont {Francica}, \citenamefont {Galve}, \citenamefont
  {Gullo},\ and\ \citenamefont {Zambrini}}]{Plastina2014}%
  \BibitemOpen
  \bibfield  {author} {\bibinfo {author} {\bibfnamefont {F.}~\bibnamefont
  {Plastina}}, \bibinfo {author} {\bibfnamefont {A.}~\bibnamefont {Alecce}},
  \bibinfo {author} {\bibfnamefont {T.~J.}\ \bibnamefont {Apollaro}}, \bibinfo
  {author} {\bibfnamefont {G.}~\bibnamefont {Falcone}}, \bibinfo {author}
  {\bibfnamefont {G.}~\bibnamefont {Francica}}, \bibinfo {author}
  {\bibfnamefont {F.}~\bibnamefont {Galve}}, \bibinfo {author} {\bibfnamefont
  {N.~L.}\ \bibnamefont {Gullo}},\ and\ \bibinfo {author} {\bibfnamefont
  {R.}~\bibnamefont {Zambrini}},\ }\href@noop {} {\bibfield  {journal}
  {\bibinfo  {journal} {Physical review letters}\ }\textbf {\bibinfo {volume}
  {113}},\ \bibinfo {pages} {260601} (\bibinfo {year} {2014})}\BibitemShut
  {NoStop}%
\end{thebibliography}%

\end{document}